\documentclass{epac98}
\usepackage{epsfig}

\newcommand{\ff}{\mbox{$\cal F\:\!$1}}

\newcommand{\mum}{\mbox{$\mu$m}}
\newcommand{\mus}{\mbox{$\mu$s}}

\title{\ff\ - An Eight Channel Time-to-Digital Converter Chip \\
      for High Rate Experiments \vspace*{-0.2cm}}

\author{G.~Braun,
\underline{H.~Fischer}$^\ast$, J.~Franz, A.~Gr\"unemaier, F.H.~Heinsius,
L.Hennig, K.~K\"onigsmann, \\[-0.05cm] 
M.~Niebuhr, M.~Schierloh, T.~Schmidt, H.~Schmitt, H.J.~Urban \\[0.1cm]
 Universit\"at Freiburg, \\[-0.05cm]
Fakult\"at f\"ur Physik, Hermann-Herder-Str. 3,
D-79104 Freiburg, Germany\\
{\tt \normalsize $^\ast$EMAIL: Horst.Fischer@cern.ch} 
\\[0.1cm]
Contribution to: \\[-0.05cm] {\it Fifth Workshop on Electronics for 
LHC Experiments - LEB99} \\[-0.05cm] Snowmass, September 20-24,  1999. 
}

\begin{document}

\maketitle

\abstract{ A new TDC chip has been developed for the COMPASS experiment
at CERN. The resulting  ASIC offers an unprecedented degree of flexibility
and
functionality. Its capability to handle highest hit and trigger input
rates as well as its  low power consumption makes it an ideal tool for
future collider and fixed target experiments. First front-end boards
equipped with the \ff\ chip have been used recently at testbeam
experiments at CERN.
A functional description
and specification for this new TDC chip is presented. }

\section{Introduction}

The COMPASS experiment at CERN was proposed to and  approved by
the CERN authorities to investigate the complex hadron structure.
One central issue of the experimental effort will be the measurement of
the contribution of gluons to the nucleon spin. Previous experiments at
CERN, DESY and SLAC made decisive advances
towards the understanding of the nucleon spin in terms of its quark
constituents, but the role of the gluon still needs more clarification.
To reach this objective a demanding state-of-the-art double-stage
spectrometer with large geometrical and dynamical acceptance will be
set-up and commissioned through 2000.

The low cross sections of the studied interactions which will provide
insight to the physics objectives of COMPASS
necessitate high beam rates. The central
detector has to be capable
to digest the high particle flux from the traversing beam at
intensities of $2\cdot 10^8$\
particles/spill. This leads consequently to the requirement of negligible
dead time for the digitization units in the data acquisition 
system of the  experiment. The anticipated
amount of data on the order of several Gigabytes per spill (2.4 seconds)
is at the
edge of todays digitisation and bandwidth technologies.

\begin{figure}[htb]
\centering
\epsfig{file=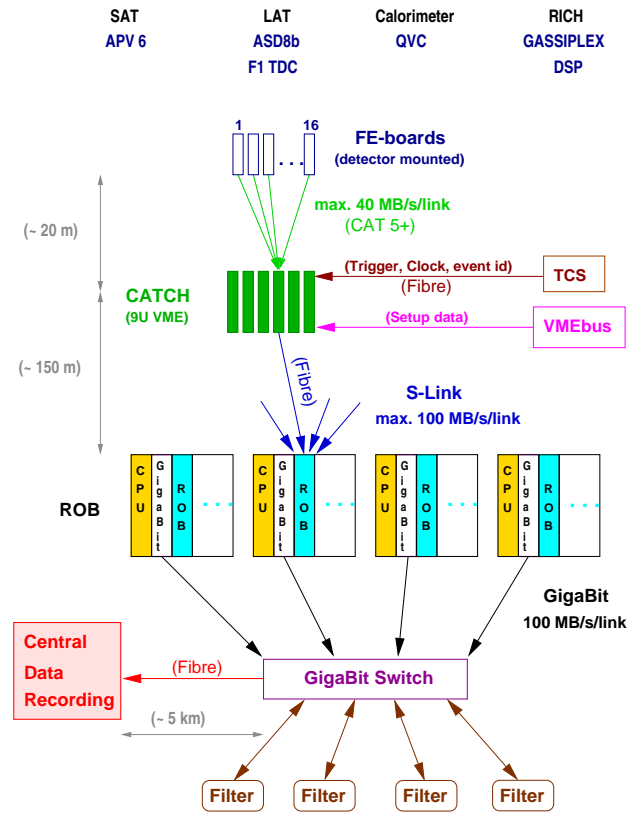, width=82.5mm}
\caption{The architecture of the COMPASS detector readout.}
\label{fig:architecture}
\end{figure}

Wherever applicable  in COMPASS measurements are digitized at the
detector by `intelligent'
front-end electronics \cite{leb98}. 
The advantages of this scheme are obvious: besides
the enormous cost savings due to obsolete data transfer and delay cables
the fact of lower noise, higher bandwidth and the possibilities for data
pre-processing open up new prospects for efficient data digitisation.
In case of analog readout a pedestal subtraction and zero suppression is
performed by the front-end devices at the COMPASS detector. To
suppress background in time measurement components only those hits are
read out which are time correlated to the trigger time. The distributed
readout architecture of the COMPASS experiment is summarized in
Figure~\ref{fig:architecture}.

The digitized data are transmitted via CAT~5+ patch cables \cite{sutp}
at rates of up to $40 \mathrm{MByte/s}$\
from the front-end to FPGA based CATCH
modules (CATCH = \underline{C}OMPASS  \underline{A}ccumulate,
\underline{T}ransfer and  \underline{C}ontrol  \underline{H}ardware) which
serve as readout drivers.
The transmission  between the front-end  and  CATCH  modules  is
based on the  serial protocol generated by the HOTLink chip~\cite{HOTLink}.
The CATCH functions as a data concentrator and a
pre-event builder as well as a control
unit of data integrity of events sent from the
front-end devices. Furthermore, the CATCH takes over automatic front-end
initialisation and calibration tasks. The same module serves also as
a remote fan-out for the COMPASS trigger distribution and time
synchronisation system (TCS).

In the next hierarchical step of the data acquisition system the data are
transmitted from the CATCH to large readout buffers (ROB) via
S-link~\cite{slink} connections. The ROBs are capable to store at
least all data
which have been collected during one spill. Again pre-event building and
checks for data consistency are performed before the sub-events are
transmitted via Gigabit Ethernet to filter computers. Here the final
event-building is performed and complete events may be rejected based on
physics constraints. Finally, the events will
be transferred at a continuous rate of up to 35 MByte/s to the central data
recording facilities at CERN where they are stored in object oriented
federated databases for later off-line analysis.

\section{The \ff\ Time Converter Core}

\begin{figure}[htb]
\centering \epsfig{file=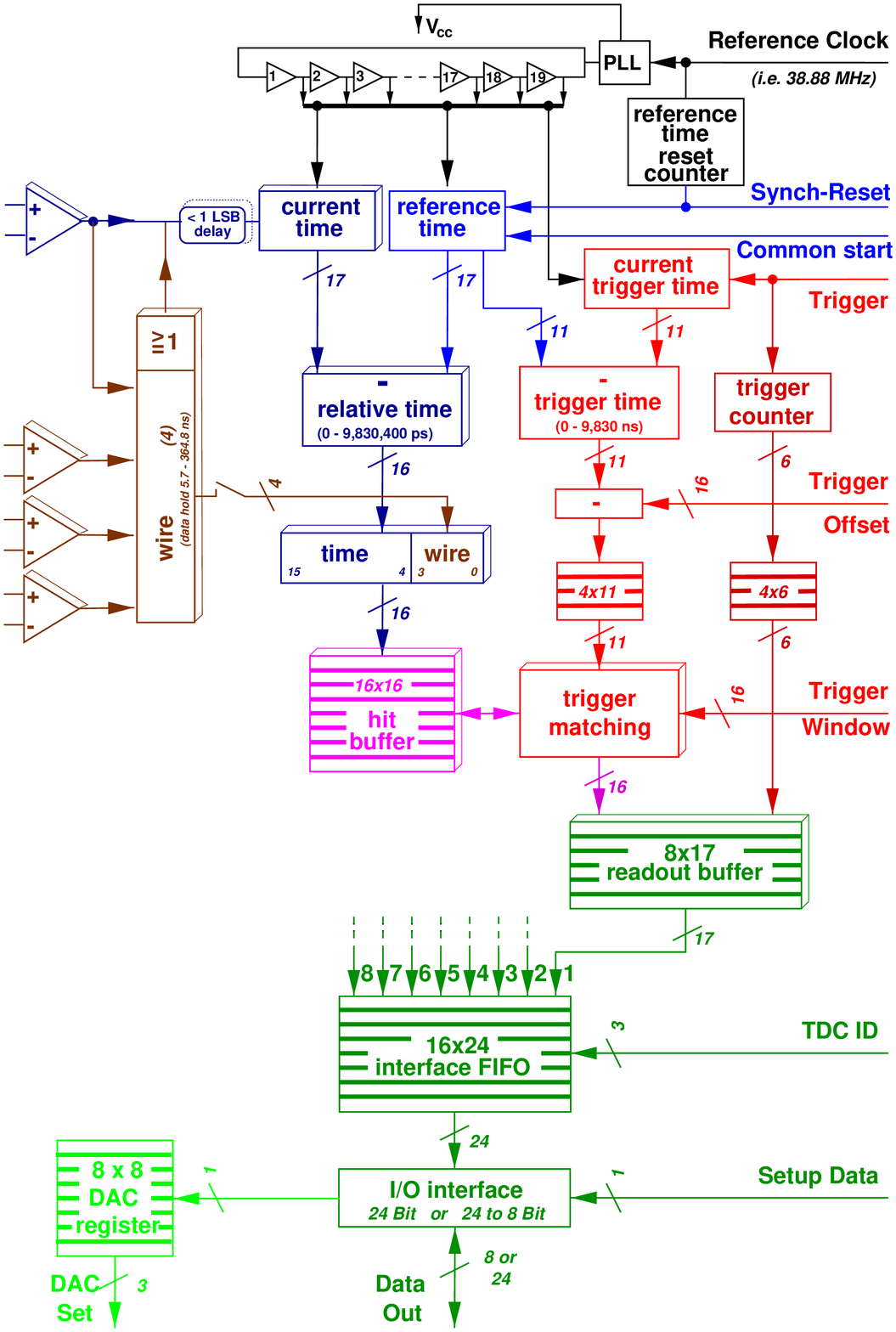, width=82.5mm}
\caption{Block
diagram for the \ff\ TDC chip. Components which are displayed as shaded
boxes are implemented for each channel, whereas non-shaded boxes are
unique and shared by all eight channels. } \label{fig:blockdia}
\end{figure}

Handling extremely high hit and data rates are the key feature of the
COMPASS detector. Hitherto existing time converter chips were not designed to
digitise six million or more hits per second with a precision of better
then $100 \mathrm{ps}$. In addition, this demanding task even has to be
performed without  any dead time during time conversion.
Given the complexity of the task to design a new TDC chip
which fulfils all COMPASS requirements the most evident
approach was to develop the \ff\ chip as ASIC
together with an industrial partner
of outstanding experience and knowledge. The outcome of the fruitful
collaboration with {\em acam-messelectronic gmbh} \cite{acam}
is a state-of-the-art multi-purpose TDC chip
with many different modes of operation in terms of functionality
and resolution. To
guide the reader through the technical description a schematic block
diagram of the chip is given in Fig.~\ref{fig:blockdia}.
The technical specifications for the TDC are summarized
in Table~\ref{techlist}.
The logic structure of the TDC architecture can be divided into the time
measurement unit, the arithmetic logic executing the trigger correlated
hit selection as well as  the readout  buffer and output interface.

\begin{figure}[htb]
\centering \epsfig{file=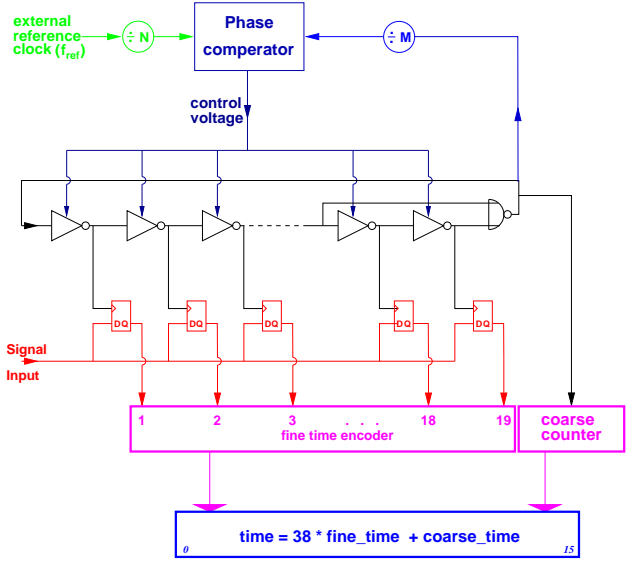, width=82.5mm}
\caption{An asynchronous
ring oscillator, stabilised by a phase locked loop, is used in
combination with a coarse counter to form the time measurement unit of
the circuit.} \label{fig:pll}
\end{figure}

The clockwork  of the \ff\ is an asynchronous ring oscillator consisting
of 19 identical voltage controlled  delay elements (see
Fig.~\ref{fig:pll}).
The dynamic range of the delay chain is extended to 16~bits by a coarse
counter. A time measurement with the \ff\ contains the combination of the
status from the array of the delay chain and the coarse time counter. The
leading and/or trailing edges of hits and the
leading edges of trigger signals
are sampled by this device without any dead time.

The process technology of 0.6 \mum\ sea-of-gates has a typical gate delay
on the order of 100 to 150~ps depending on the batch of the production
run, the ambient temperature and the supply voltage of the delay
elements. To maintain stable gate propagation delays over extended
periods  a phase locked loop (PLL) is used to control the ring oscillator
frequency. Thus the time measurement circuit maintains a constant time
resolution and is tolerant to very large variations in temperature $(\pm
75^\circ \mathrm{C})$. The
frequency of the ring oscillator and the time resolution of the \ff,
respectively, can be selected with two pre-scalers $M$\ and $N$ (see
Fig.~\ref{fig:pll}). The
ratio between the pre-scalers for the reference clock $N$\ and the ring
oscillator $M$\ determines the cycle time of the asynchronous ring
oscillator, which is 152 times the propagation time through a single
delay element. Therefore, the user can chose the external
reference clock freely  between 500~kHz and 40~MHz.

Unlike a delay locked loop (DLL), a PLL filters any jitter of the
reference clock signal. Therefore, no special precaution for the
clock distribution system is necessary to provide  the uncompromised
high resolution of the TDC even in distributed  applications.
A carefully chosen floor plan together with the introduction of
several separate power and ground supplies for the pad ring, high current
output lines, and the time measurement unit are used to
minimise the noise coupling.

The signal propagation of each input stage can be delayed by
approximately 1 LSB divided in 64 individual steps.
This feature can be used to improve the resolution of the TDC
by a factor of two. In case the \ff\ is operated in the
high resolution mode
two neighboured input channels are connected internally and hence the
original signal is split into two channels.
The delay  of one of the two signals is chosen such that it
is half a LSB relative to its sister channel. This way additional phase
information on the signal can be derived and be used to improve the
time measurement to a resolution of $30 \, \mathrm{ps \; RMS}$.
In this mode the dynamic range of the TDC is reduced by a
factor of two.

\section{The \ff\ as Latch Device}
Another mode how to operate the \ff -chip is the latch or hit mode.
This mode  is suitable  for the readout
of detectors which do not require precise time information for event
pattern reconstruction but only  time stamps for event building in a
pipelined data acquisition system or for background suppression.
Therefore it
matches the requirements for MWPC readout or standard latch units and can
be regarded as a cost efficient replacement of existing commercial
products.

In the latch mode the number of input channels of the \ff\ is increased
to 32. Each group of four of the 32 channels is connected to a fourfold
input register and a logic OR which is linked to one of the eight time
measurement channels of the \ff\ (see Fig. \ref{fig:blockdia}). When a hit
arrives on one of four combined channels the next clock cycle of the
asymmetric ring oscillator starts a $6\; \mathrm{bit}$\ counter. This
counter, which defines the strobe length during which the registers
accept input signals, is synchronous to the coarse counter, thus running
with a period of $\ff_{LSB} \times 38 \approx 5.7 \; \mathrm{ns} $, where
$\ff_{LSB} = 150 \mathrm{ps}$\ refers to a typical digitisation
bin size of the
\ff\ in the standard resolution mode. After a pre-set time of
$5.7 \; \mathrm{ns} <t_{strobe} <  364.8 \; \mathrm{ns} \
(64 \times 5.7 \; \mathrm{ns})$\
the four inputs are switched to a second hit input
register and a time stamp is taken with the full accuracy of the
standard mode. The 12 most significant bits of the time stamp are placed
in front of the four bits from the input register and transferred to the
hit buffer. To ensure that no hits are lost while the
input to the hit registers are switched  both hit registers will accept
signals during  an overlap of about $2\; \mathrm{ns}$. Although it is
desirable to have the overlap as short as possible small variations in
production processes and the danger of efficiency gaps may require a
longer overlap.

\section{Data Sparsification}

Digitized time stamps from hits are stored in dual port buffers.
Each \ff\ channel has its individual buffer which can store up to
16 hits before it blocks further data input. The hit buffers are addressed
in random access mode such that the  trigger matching unit can search
for data which belong to a coincident trigger.
In the high resolution mode the hits are stored alternately in both
hit buffers of the two channels. The
procedure to store the data alternately permits
higher data rates or longer trigger latencies because it is equivalent to
double a single hit buffer to 32 words. A second advantage is the
improved performance of the trigger matching unit since two units work in
parallel on the data set in the hit buffers.

\begin{figure}[htp]
\centering \epsfig{file=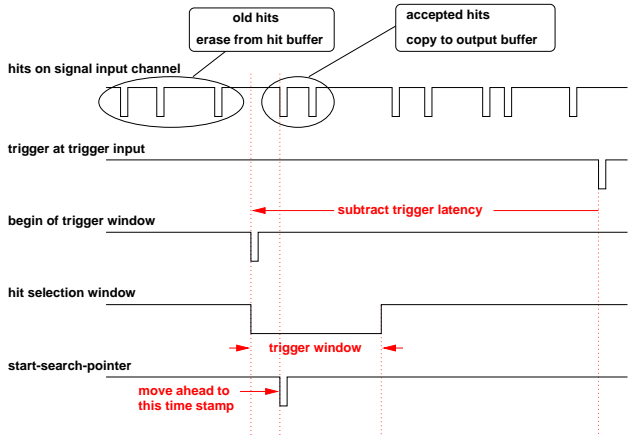, width=82.5mm}
\caption{\small Example for the trigger matching procedure.}
\label{fig:trigger-select}
\end{figure}

An arriving trigger signal is digitized by the same time measurement unit
as the hit-signals. A programmable trigger latency time is subtracted
from the measured trigger time to account for the time needed to form
and to distribute the trigger to the distant front-ends.
Trigger time-stamps are stored in a common four word deep FIFO until the
arithmetic logic units
are available to start with the trigger matching process.
A programmable trigger window defined by the maximum time spread of a
detector, e.g. drift time in wire chambers or time of flight, is loaded
at the initialisation of the chip. A hit is considered to coincide with
a trigger if its time stamp is within the latency corrected trigger
time and the upper limit of the trigger window
(see Fig.~\ref{fig:trigger-select}).
Local bookkeeping of all triggers arriving at the \ff\  trigger input 
is handled by a  six bit counter.
In the  case of a trigger buffer overflow
the number of missed trigger signals
can be reconstructed from a comparison of the trigger counter and
from the trigger time stamps.

The search mechanism which is used in the
trigger matching unit uses two independent pointers.
The read pointer specifies the memory address currently being accessed
to look for a hit which matches the trigger time.
The start-search pointer marks the address where the search is
supposed to start when the next trigger is loaded from the trigger FIFO.
The start-search pointer is set to the read pointer position
at the location of the first hit which matches the trigger time.
All hits previous to this particular hit are deleted because
they are
outside of the future regions of interest. The selected hits are copied to
the readout buffer.
The search for further hits which belong to the same event is continued
until either no more hits are stored or the next found time stamp in the
hit buffer is younger than the region of the trigger window allows.
Regions of interest for a sequence of triggers can overlap for those
detectors which require long trigger windows. Therefore,
selected hits are not deleted from the hit buffer when copied to
the readout buffer.

When  no triggers arrive for longer periods the write pointer may
catch up with the start-search pointer. In this case new hits would not be
accepted by the hit buffer anymore and the time stamps would be
lost until the next
trigger starts the clean-up process of the hit buffer.
To avoid this,
an artificial fake trigger is generated internally at regular intervals
as soon as the trigger FIFO is empty.
The fake trigger is handled like any
real trigger except that no data are copied to the readout buffer.
The generation of fake triggers helps to clean-up the hit
buffer from old hits and guarantees unambiguous time measurements.

\section{Readout}

The introduction of  dedicated readout buffers permits  fast data readout
while hits belonging to the next trigger are already selected in parallel
by the trigger matching units.
When the trigger matching process for a particular event is completed
on all channels of a \ff\ the data ready signal is set and the output
interface is prepared for readout. In case several chips are connected
to a single front-end data bus the readout is controlled by a circulating
token. The token control in the \ff\ has been optimised to avoid wait states
when a token is handed over to the next chip.

The high speed readout interface of the \ff\ can be clocked with
a frequency of up to $50 \mathrm{MHz}$. Two different modes for
data readout are supported by the \ff: 8 bit and 24 bit readout.
The eight bit readout is used in the COMPASS experiment for  applications
where serial data interfaces and serial data links are used. The eight
bit readout mode accords with
the timing specifications of the
HOTLink parallel to serial converter chip~\cite{HOTLink}.
For a safe timing the \ff\
provides a bus-write-enable and a delayed bus clock with an adjustable
skew for the HOTLink.

The 24 bit parallel readout can be used where highest data rates are
anticipated and no data serializer is used. In this mode the user can
exploit the full data throughput capabilities of the TDC chips.

\section{TDC Initialisation}

All data communications between the CATCH boards and the TDC front-end
boards are exclusively serial. For the setup procedure, the TDC has a TTL
serial data input which can be operated at a sustained
rate of $ 10 \; \mathrm{MBit/s} $. The TDC serial interface is designed
as a  fourfold oversampling device.
For different applications either the PLL reference  clock or an optional
dedicated setup clock can be used.
The format for the setup data
is characterised by two start and two stop
bits, three bits for the chip address,
a common address bit and four bits for
register addresses followed by 16~bits for setup data.
Since some data may be common  for all chips on a
board, these can be loaded in parallel by making use of the common
address bit. In case of errors in the start or stop bits,
the data reception is refused by the \ff.

\section{DAC Interface}

In the COMPASS experiment the AD8842 octal 8-Bit digital-to-analog
converter (DAC)~\cite{dac} is used to control the thresholds
of discriminators
of some detectors. The AD8842 provides eight general purpose
digitally controlled voltage converters with separate voltage inputs.
Each AD8842 has its own register that holds the output state. The
registers are updated from a 3-wire serial input digital interface.

The \ff\ is used to interface the serial setup lines from the CATCH
to the DAC and to store the threshold values for multiple down-loads.
 The \ff\ contains eight registers of 1 Byte 
which may hold all register values for one AD8842.
The  data can be formatted
and send via the 3-wire output interface to the AD8842.
This operation does not interfere with normal TDC time conversion.

\begin{table}
\caption{ Technical description of the \ff. }
\label{techlist} \medskip \begin{center}
\begin{tabular}{l} {\bf Number of channels:}  \\ \hspace*{0.2cm} 4
for high resolution mode \\ \hspace*{0.2cm} 8 for standard
resolution mode  \\ \hspace*{0.2cm}32 for latch mode
\\[3pt] {\bf Digitisation bin size:}  \\ \hspace*{0.2cm}75 ps for high
resolution mode \\ \hspace*{0.2cm}150 ps for standard operation
mode \\ \hspace*{0.2cm}5700 ps for latch mode\\
\hspace*{0.2cm} (typical values)
\\[3pt]
{\bf Reference-clock frequency:  }      \\
\hspace*{0.2cm}Between 500 kHz and 40 MHz  \\
\hspace*{0.2cm}(Clock is used for self calibration only) \\[3pt]
{\bf Differential non-linearity: } \\ 
\hspace*{0.2cm}Less than 0.3 LSB \\[3pt] 
{\bf Integral non-linearity: }   \\
\hspace*{0.2cm}Less than one time bin\\[3pt] {\bf Variation with
temperature or supply  voltage: } \\ \hspace*{0.2cm}Less than one time bin \\[3pt]
{\bf Dynamic range:  }     \\ \hspace*{0.2cm}16 bits \\[3pt] {\bf
Double pulse resolution:}  \\ \hspace*{0.2cm}Typical 22 ns \\[3pt]
{\bf Digitisation and readout dead time:}  \\ \hspace*{0.2cm}None
\\[3pt] {\bf Hit buffer size: }   \\ \hspace*{0.2cm}32
measurements for high resolution mode \\ \hspace*{0.2cm}16
measurements for standard mode \\ \hspace*{0.2cm}16 measurements
for latch mode\\[3pt] {\bf Output buffer size:}  \\
\hspace*{0.2cm}8 measurements  \\[3pt] {\bf Common readout interface: }
\\ \hspace*{0.2cm}16 measurements  \\[3pt]
{\bf Trigger buffer size: }   \\ \hspace*{0.2cm}4 \\[3pt]
{\bf Trigger latency: }   \\  \hspace*{0.2cm} up to full dynamic
range of digitisation unit \\
\hspace*{0.2cm} (typical values: 9.8\mus , 4.9 \mus ,
standard, \\ \hspace*{0.2cm} high resolution mode, respectively) \\[3pt]
{\bf Power supply: } \\ \hspace*{0.2cm}5.0 V \\[3pt]
{\bf Power consumption: } \\ \hspace*{0.2cm}80 mA \\[3pt]
{\bf Temperature range:  }  \\ \hspace*{0.2cm}-40 to  +85 degree
centigrade \\[3pt] {\bf Hit input: } \\ \hspace*{0.2cm}LVDS, LVPECL or TTL
\\[3pt] {\bf Package:} \\ \hspace*{0.2cm}160 PQFP \\[3pt]
\end{tabular}
\end{center}
\end{table}

\begin{figure}[htp]
\centering \epsfig{file=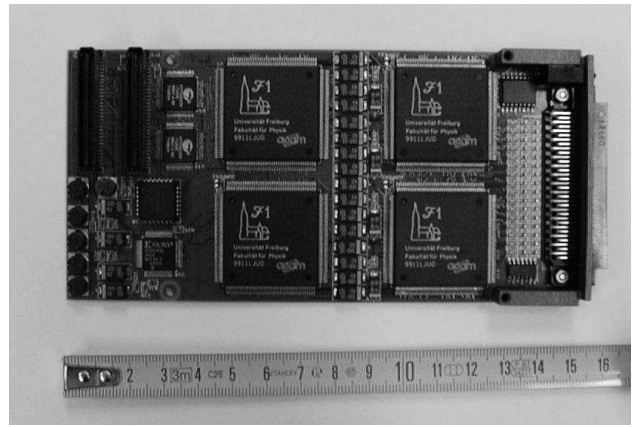, width=82.5mm}
\caption{\small This picture shows a 32 channel TDC board.
The board dimensions and board interface connectors are
 according to the CMC standard~\cite{mezzanine}.}
\label{fig:tdc-cmc}
\end{figure}

\newpage
\section{Acknowledgement}

Our grateful appreciation goes to the local electronic workshop
and engineer team, whose support during the development and
production phase of the electronic components was essential.
We also recognise the endeavours of our collaborators from the COMPASS
collaboration and by the staffs of the collaborating institutions during
several test periods at CERN. In particular we deeply appreciate the many
stimulating discussions of our colleagues involved in front-end
electronics development. The developments described in this report are
supported by the German Bundesministerium f\"ur Bildung, Wissenschaft,
Forschung und Technologie.


\end{document}